\documentclass[aps,prd,reprint,groupedaddress]{revtex4-2}
\usepackage{amsmath,amsfonts}
\usepackage{enumitem}
\usepackage{xcolor}

\DeclareMathSymbol{\shortminus}{\mathbin}{AMSa}{"39}
\newcommand{\PS}{Pol\'a\v{c}ek-Siegel }
\newcommand{\gGS}{generalized Green-Schwarz }
\newcommand{\GPS}{G_{\mathrm{PS}}}
\newcommand{\GS}{G_{\mathrm{S}}}

\newcommand{\aL}{\overline{a}}
\newcommand{\bL}{\overline{b}}
\newcommand{\aR}{\underline{a}}
\newcommand{\bR}{\underline{b}}
\newcommand{\cl}{\overline{c}}
\newcommand{\Cr}{\underline{c}}
\newcommand{\dl}{\overline{d}}
\newcommand{\dr}{\underline{d}}
\newcommand{\el}{\overline{e}}
\newcommand{\er}{\underline{e}}
\newcommand{\fl}{\overline{f}}
\newcommand{\fr}{\underline{f}}
\newcommand{\gl}{\overline{g}}

\newcommand{\alphaL}{\overline{\alpha}}
\newcommand{\betaL}{\overline{\beta}}
\newcommand{\gammaL}{\overline{\gamma}}
\newcommand{\alphaR}{\underline{\alpha}}
\newcommand{\betaR}{\underline{\beta}}
\newcommand{\gammaR}{\underline{\gamma}}

\begin{document}
\title{All-order generalized Green-Schwarz transformations}

\author{Achilleas Gitsis}
\email[]{achilleas.gitsis@uwr.edu.pl}
\author{Falk Hassler}
\email[]{falk.hassler@uwr.edu.pl}
\homepage{https://www.fhassler.de}
\affiliation{University of Wroc\l{}aw, Faculty of Physics and Astronomy, Maksa Borna 9, 50-204 Wroclaw, Poland}

\date{\today}

\begin{abstract}
	Compatibility with T-duality severely constrains higher-derivative corrections to the low-energy supergravity limits of string theory. For example, it suggests that Lorentz transformations for heterotic strings are modified in precisely the way required for the Green-Schwarz anomaly cancellation mechanism. A systematic procedure to construct the resulting generalized Green-Schwarz transformations is the generalized Bergshoeff-de Roo identification (gBdRi) \cite{Baron:2018lve,*Baron:2020xel}. Although it in principle allows computing $\alpha'$-corrections to higher and higher orders, technically it becomes unfeasible beyond $\alpha'^2$. We revisit this problem with an alternative approach to the gBdRi, which we have recently developed. It gives rise to a very simple all-order transformation law whose closure we verify by explicitly computing the resulting gauge algebra.
\end{abstract}


\maketitle

\section{Introduction}
Higher-derivative corrections of the low-energy, supergravity limit of string theory have a wide range of applications spanning from string phenomenology and cosmology, over the holographic analysis of conformal field theories through the AdS/CFT correspondence, to the study of black holes and their entropy. Obtaining them directly from string amplitudes or higher-loop $\beta$-functions in non-linear $\sigma$-models is technically very challenging. Hence, only the leading order corrections are known in full generality. One of the main challenges is that the number of terms which are compatible with the diffeomorphisms and gauge field transformations grows extremely fast with an increasing number of derivatives. Consequentially, a lot of work is needed to fix the respective coefficients in the effective action by comparing them with results from string amplitudes or $\beta$-functions. A possible way to remedy this problem is to find additional symmetries which are more restrictive and thereby significantly reduce the admissible terms in the effective action. Unfortunately, these symmetries are not manifest and thus realized in a convoluted, non-linear way.

Supersymmetry is the most obvious example. But due to longstanding challenges in finding the relevant corrected transformations and their invariants, a considerable amount of work has been spent to use duality symmetries as an alternative. After the reduction on an $n$-dimensional torus where only massless modes are kept, a global O($n$,$n$) symmetry emerges \cite{Meissner:1991zj,*Meissner:1991ge,Sen:1991zi,Meissner:1996sa}. This symmetry captures constant diffeomorphisms, $B$-field transformations and T-dualities on the torus used in the reduction. Because it is much more constraining than diffeomorphisms and $B$-field transformations alone, it drastically reduces the number of allowed terms in the effective action. Eventually, it only leaves very few coefficients which still have to be fixed by comparing with amplitudes or $\beta$-functions. Despite this success, the explicit realization of the pivotal O($n$,$n$) duality symmetry after the reductions is very complicated and has to be painstakingly computed order by order \cite{Codina:2021cxh,Garousi:2023kxw,Wulff:2024mgu,Ameri:2025bei}.

Hence, a lot of effort has been invested to make this symmetry manifest even before the reduction in the framework of double field theory \cite{Siegel:1993th,Hull:2009mi,*Hohm:2010pp}, and the closely related generalized geometry \cite{Hitchin:2003cxu,Gualtieri:2003dx}, where the metric and the $B$-field are unified to form the generalized metric,
\begin{equation}
	\label{eqn:genMet}
	H_{I J} = \begin{pmatrix}
		g_{i j} - B_{i k}g^{k l}B_{l j} &  & B_{i k}g^{k j} \\
		-g^{i k}B_{k j}                 &  & g^{i j}
	\end{pmatrix}\,,
\end{equation}
an element of the coset O($d$,$d$)/O($d-1$,1)$\times$O(1,$d-1$). It arises from an unconstrained duality group element called the generalized frame $E^A{}_I$ by
\begin{equation}
	H_{I J} = E^A{}_I H_{AB} E^B{}_J \,, \quad
	\text{with} \quad H_{AB} = \begin{pmatrix}
		\eta_{\aL\bL} & 0               \\
		0             & - \eta_{\aR\bR}
	\end{pmatrix}\,.
\end{equation}
Here $\eta_{\aL\bL}$ is the Minkowski metric corresponding to the first O($d-1$,1) factor in the coset, while $\eta_{\aR\bR}$ governs the second one. We will always distinguish between the string's left- and right-moving sector by using over- and under-barred indices like here. Diffeomorphisms and $B$-field transformations are mediated by the generalized Lie derivative
\begin{equation}
	\label{eqn:genLie}
	\mathcal{L}_{\xi} E^A{}_I = \xi^J \partial_J E^A{}_I -(\partial^J \xi_I - \partial_I \xi^J) E^A{}_J\,.
\end{equation}
Keep in mind that indices are raised and lowered with the invariant $\eta$-metric of the duality group,
\begin{equation}
	\eta_{IJ} = \begin{pmatrix} 0          & \delta_i^j \\
                \delta_j^i & 0
	\end{pmatrix}
	\quad\text{or}\quad
	\eta_{AB} = \begin{pmatrix} \eta_{\aL\bL} & 0             \\
                0             & \eta_{\aR\bR}
	\end{pmatrix}\,,
\end{equation}
for either curved, like $I$, $J$, \dots, or flat indices, like $A$, $B$, \dots\,. Moreover, we use the partial derivative $\partial_I = \begin{pmatrix} \partial_i & 0 \end{pmatrix}$ and thereby solve the section condition explicitly. Working with the frame $E_A{}^I$ becomes mandatory if we want to find a background independent formulation of $\alpha'$-corrections \cite{Marques:2015vua,Hohm:2016yvc}. It requires a second local symmetry which is known as double Lorentz symmetry \cite{Geissbuhler:2013uka}. As the name suggests, it is generated by the transformations with the parameter $\Lambda\in\mathfrak{o}$($d-1$,1)$\times\mathfrak{o}$(1,$d-1$) that act on the frame by
\begin{equation}\label{eqn:doublelorentz}
	\delta_\Lambda E_A{}^I = \Lambda_A{}^B E_B{}^I
	\quad\text{with}\quad
	\Lambda_{AB} = \begin{pmatrix} \Lambda_{\aL\bL} & 0                \\
                0                & \Lambda_{\aR\bR}
	\end{pmatrix}\,.
\end{equation}

In this article, we discuss how the double Lorentz transformations \eqref{eqn:doublelorentz} can be modified to take higher-derivative corrections into account. The latter can only appear here, because neither diffeomorphisms nor the $B$-field transformations will change -- and with them the generalized Lie derivative \eqref{eqn:genLie}. Figuring out the explicit form of admissible corrections is subtle, because $\delta_\Lambda E_A{}^I$ is highly constrained by requiring that all possible transformations close into a Lie algebra. The seminal work \cite{Marques:2015vua} reveals that a consistent modification which contains derivatives arises after complementing \eqref{eqn:doublelorentz} with the additional transformation
\begin{equation}
	\label{eqn:gGS1}
	\delta_\Lambda E_{\underline{a}\overline{b}} =\frac{a}{2}D_{\underline{a}}\Lambda^{\underline{c}\underline{d}}F_{\overline{b}\underline{c}\underline{d}} + \frac{b}{2}D_{\overline{b}}\Lambda^{\overline{c}\overline{d}}F_{\underline{a}\overline{c}\overline{d}}\,.
\end{equation}
It contains the two free parameter $a$, $b$, and is written exclusively in terms of flat indices by using
\begin{equation}
	\delta_\Lambda E_{AB} : = \delta_\Lambda E_A{}^I E_{BI} = \begin{pmatrix}
		\Lambda_{\aL\bL}          & - \delta_\Lambda E_{\aR\bL} \\
		\delta_\Lambda E_{\aR\bL} & \Lambda_{\aR\bR}
	\end{pmatrix}\,.
\end{equation}
Both diagonal elements are already fixed at the leading order in \eqref{eqn:doublelorentz} and do not receive any corrections. Hence, two-derivative corrections are controlled by the off-diagonal elements given by \eqref{eqn:gGS1} in terms of the flat derivative
\begin{equation}
	D_A = E_A{}^I \partial_I
\end{equation}
and the generalized fluxes
\begin{equation}
	F_{ABC} = 3 D_{[A} E_B{}^I E_{C]I}\,.
\end{equation}
These new transformations are known under the name generalized Green-Schwarz transformations (gGSt) because they contain a correction to Lorentz transformations which is crucial for the Green-Schwarz anomaly cancellation mechanism \cite{Green:1984sg}. After imposing that the four-derivative effective action is invariant under this deformed symmetry,  it is fixed completely up to a global coefficient. After fixing $a$=$-\alpha'$, $b$=0 (heterotic) or $a$=$b$=$\alpha'$ (bosonic), one recovers the correct low-energy effective actions of the respective string theories \cite{Marques:2015vua}.

Because the transformations \eqref{eqn:gGS1} only close after neglecting terms with more than two derivatives, they need to be further modified with terms carrying four and more derivatives. These terms can be systematically constructed with a technique known as the generalized Bergshoeff-de Roo identification (gBdRi) \cite{Baron:2018lve,Baron:2020xel}. For up to four derivatives this has been done first for just the parameter $b$ \cite{Baron:2018lve} and later for both parameters \cite{Baron:2020xel}. As before, the arising symmetry fixes the action at order $\alpha'^2$ up to an overall coefficient and reproduces the expected results from the bosonic and heterotic string \cite{Hronek:2022dyr}. Unfortunately the gBdRi becomes quickly cumbersome with an increasing number of derivatives. Therefore, it is not feasible to use it beyond the order $\alpha'^2$.

Recently, we presented the twisted \PS construction \cite{Gitsis:2024gfb} as an alternative approach. Following an idea initiated in \cite{Polacek:2013nla} and further developed in \cite{Butter:2022iza}, it utilizes an extended space (the mega-space) to construct covariant tensors under generalized diffeomorphisms which are generated by the generalized Lie derivative \eqref{eqn:genLie}, and double Lorentz transformations. After an appropriate twist, it gives a more geometric perspective on the gBdRi. Moreover, it allows presenting the gGSt in an intermediate form which is much simpler to handle than the final expression. In the following, we make use of these new insights to gGSt to all orders in $\alpha'$ and to obtain the algebra they close into.

\section{All-order \gGS transformations} \label{sec:GS}
\subsection{Twisted \PS construction} \label{TPS}
To compute the gGSt to all orders in $\alpha'$, we will employ the twisted \PS construction presented in \cite{Hassler:2024yis}. One of its most important features is that the duality group O($d$,$d$) is extended to the group $\GPS$ which is generated by three types of generators, $K_{AB}$, $R^A_\alpha$, and $R_{\alpha\beta}$, with the non-trivial commutators
\begin{equation}
	\label{eqn: coms}
	\begin{aligned}
		\relax[K_{AB}, K_{C D}]             & = \phantom{-} 2\eta_{[A|[C}K_{D]|B]}\, ,                        \\
		[R^A_{\alpha}, R^B_{\beta}]         & = - 2 \kappa_{\alpha\beta}K^{AB} + \eta^{AB}R_{\alpha\beta}\, , \\
		[K_{A B}, R^C_{\gamma}]             & = -\delta_{[A}^C \eta_{B]D}R^D_{\gamma}\, ,                     \\
		[R^A_{\alpha}, R_{\beta\gamma}]     & = -2\kappa_{\alpha[\beta}R^A_{\gamma]}\, , \qquad \text{and}    \\
		[R_{\alpha\beta}, R_{\gamma\delta}] & = -4\kappa_{[a|[\gamma}R_{\delta]|\beta]}\, .
	\end{aligned}
\end{equation}
In addition to the invariant metric $\eta_{AB}$ of O($d$,$d$), we need a second non-degenerate metric $\kappa_{\alpha\beta}$. It has been discussed in detail in \cite{Hassler:2024yis} but for most of the discussions here, we do not need its explicit form. Both metrics are used to raise and lower the respective indices. The reason why $\GPS$ appears is that in addition to the O($d$,$d$)-valued generalized frame field $E_A{}^I$, there are also various connections which play the role of a suitable generalized analogue of the spin connection known from differential geometry. All of them are combined into the frame field,
\begin{equation}
	\mathcal{E} = \mathcal{A} E \in \GPS\,,
\end{equation}
on the mega-space.

Besides $\GPS$, there is also the generalized structure group $\GS\subset\GPS$. Its action on $\mathcal{A}$ and $E$ is encoded in
\begin{equation}\label{eqn:gaugetr}
	\mathcal{A}^{-1}\delta \mathcal{A} + \delta E E^{-1} = -D_A\xi^{\alpha}R^A_{\alpha} - \xi^{\alpha} R_{\alpha} + \xi^\alpha \mathcal{A}^{-1} \tau_\alpha \mathcal{A}\,,
\end{equation}
where $R_\alpha$ is given by
\begin{equation}
	R_\alpha := \tfrac12 f_{\alpha}{}^{\beta\gamma} R_{\gamma\beta}\,.
\end{equation}
The structure coefficients $f_{\alpha\beta}{}^\gamma$ that appear here define $\GS$'s Lie algebra in terms of the generators $\tau_\alpha$ with
\begin{equation}\label{eqn:GSalgebra}
	[ \tau_\alpha, \tau_\beta ] = - f_{\alpha\beta}{}^\gamma \tau_\gamma\,.
\end{equation}
One should also note that $\kappa_{\alpha\beta}$ is a bi-invariant linear form, namely $f_{\alpha(\beta}{}^\delta \kappa_{\gamma)\delta}$=$0$, and thus turns $\GS$ into a metric Lie group. As we will work a lot with \eqref{eqn:gaugetr}, it is convenient to further simply this  expression by suppressing all indices. More precisely, we define
\begin{equation}
	\xi = \xi^\alpha \tau_\alpha \,, \quad \xi^{\alpha}R_{\alpha} = \xi_R \,, \quad \text{and} \quad
	D_A\xi^{\alpha} R^A_{\alpha} = D\xi\,.
\end{equation}
Note that the transformation \eqref{eqn:gaugetr} is easily extended to also incorporate generalized diffeomorphisms in addition to $\GS$ transformations. The full expression is presented in \cite{Gitsis:2024gfb}. However, as we will argue below, generalized diffeomorphisms decouple from the gGSt and only reappear in the closure of the gauge algebra. Hence, we will only highlight their contributions where they are relevant and otherwise set the parameter $\xi_A$ which controls them to zero.

\subsection{Partial gauge fixing}
Eventually, $\mathcal{A}$ has to be completely fixed in terms of $E$ and its derivatives. There are two mechanisms,
\begin{enumerate}
	\item\label{item:gf} partial gauge fixing, and
	\item torsion constraints,
\end{enumerate}
at play to achieve this goal. In the original gBdRi \cite{Baron:2018lve,Baron:2020xel}, both are performed at the same time in each step, but the \PS construction allows treating them independently. Hence, here we will only consider \ref{item:gf} and thereby obtain results which are valid for arbitrary choices of torsion constraints.

As an element of $\GPS$, $\mathcal{A}$ can be written as
\begin{equation}\label{eqn:expMap}
	\mathcal{A} = \exp \left( \sum_{n=1}^\infty c_n A^n \right)
\end{equation}
for any choice of the coefficients $c_n$. In principle, any choice of $c_n$ will work, as different choices can be related by field redefinitions of the generalized connections
\begin{equation}
	A = A^\alpha_B R_\alpha^B + \tfrac12 A^{\alpha\beta} R_{\beta\alpha}\,.
\end{equation}
But as we will see soon, there is a preferred choice which will lead to very simple results. To motivate it, we first need to take into account that $\GPS$ has a natural $\mathbb{Z}_2$ grading which originates from the left- and right-moving sector of the underlying closed string. It becomes manifest by bringing $\eta_{AB}$ and $\kappa_{\alpha\beta}$ into the block diagonal forms
\begin{equation}
	\eta_{AB} = \begin{pmatrix}
		\eta_{\aL\bL} & 0             \\
		0             & \eta_{\aR\bR}
	\end{pmatrix}\,, \quad \text{and} \quad
	\kappa_{\alpha\beta} = \begin{pmatrix}
		\kappa_{\alphaL\betaL} & 0                      \\
		0                      & \kappa_{\alphaR\betaR}
	\end{pmatrix}
\end{equation}
with $\eta_{\aR\bR}=-\eta_{\aL\bL}$. Under this grading $A$ for example decomposes into a (anti-)chiral part $A_+$ and a mixed chiral part $A_-$,
\begin{equation}
	A = A_+ + A_-\,,
\end{equation}
with
\begin{equation}
	A_- = A^{\alphaL}_{\bR} R_{\alphaL}^{\bR} + A^{\alphaR}_{\bL} R_{\alphaR}^{\bL} + \tfrac12 A^{\alphaL\betaR} R_{\betaL\alphaR} + \tfrac12 A^{\alphaR\betaL} R_{\betaR\alphaL}
\end{equation}
and $A_+$ respectively. Moreover, $\GS$ is purely chiral and therefore only the structure coefficients $f_{\alphaL\betaL}{}^{\gammaL}$ and $f_{\alphaR\betaR}{}^{\gammaR}$ in \eqref{eqn:GSalgebra} are non-zero.

Our gauge fixing of $A$ is
\begin{equation}\label{eqn:gaugefixing}
	A_+ = 0\,,
\end{equation}
implying that $A$ only has mixed chiral contributions which will be fixed by torsion constraints. To also keep the argument of the exponential map in \eqref{eqn:expMap} mixed chiral, all even coefficients $c_{2n}$=$0$ have to vanish. A suitable choice for the first two of all remaining coefficients has been identified in \cite{Gitsis:2024gfb} as $c_1$=$1$ and $c_3$=$1/3$ by imposing that the maximal number of terms cancels. Pushing the computation there a bit further, we found after pages of computation $c_5$=$1/5$. Hence, one might guess that
\begin{equation}
	\label{eqn:cn}
	c_{2n+1} = \frac1{2n + 1}
\end{equation}
is the best choice, and indeed, we will see that this is true. It is special in two ways: First, it allows us to write \eqref{eqn:expMap} as
\begin{equation}\label{eqn:sqrtCayley}
	\mathcal{A} = \sqrt{\frac{1+A}{1-A}}\,,
\end{equation}
and second, if we expand $\mathcal{A}$ directly to
\begin{equation}
	\mathcal{A} = \sum_{n=0}^\infty d_n A^n
\end{equation}
it is the only choice for which two adjacent coefficients
\begin{equation}\label{eqn:dn}
	d_{2n} = d_{2n+1} = \frac{(2n - 1)!!}{(2n)!!}
\end{equation}
match. Later on, we will see that this is the property which gives rise to the extensive cancellations we observed in our painstaking initial computations.

\subsection{Transformation of the generalized frame}
Due to the gauge fixing \eqref{eqn:gaugefixing}, the transformation of the generalized frame is modified and eventually becomes the gGSt we are after. Organizing \eqref{eqn:gaugetr} according to the number of derivatives which contribute to the individual terms, we find the recursive relation
\begin{equation}
	\label{eqn:gauge}
	\delta A^{(n)} + \delta E^{(n)}E^{-1} - \widetilde{\xi}^{(n)} = \left. - D\xi - \mathcal{A}^{-1}\delta \mathcal{A} + \mathcal{A}^{-1}\xi \mathcal{A} \right|^{(<n)}\,.
\end{equation}
With $A^{(n)}$ we denote all contributions to $A$ with $n$ derivatives. Moreover we use $|^{(<n)}$ to say that on the right-hand side only $A$, $\delta E E^{-1}$ and $\widetilde{\xi}$ which contain less than $n$ derivatives are considered. Also note that we have introduced here $\widetilde{\xi} = \xi - \xi_R$ to keep this and the following relations a bit shorter. The process of partial gauge fixing now works as follows: First, we seed this relation with
\begin{equation}
	\delta E^{(0)} E^{-1} = \widetilde{\xi}^{(0)} = \Lambda^{(0)}\,,
\end{equation}
where $\Lambda^{(0)}$ is the parameter for double Lorentz transformations. Moreover, we know that $\delta E^{(n)} E^{-1}$ with $n>0$ can only have mixed chiral $K_{A B}$ generators because double Lorentz transformations are modified. All other mixed chiral generators that arise from the right-hand side go to $\delta A^{(n)}$. Finally, all chiral contributions give rise to the gauge fixing of $\widetilde{\xi}^{(n)}$. For this to work, the structure group $\GS$ has to be large enough and contain all generators $R^{\aL}_{\betaL}$, $R^{\aR}_{\betaR}$, $R^{\alphaL\betaL}$, and $R^{\alphaR\betaR}$. We constructed an appropriate group in \cite{Gitsis:2024gfb} and will use it in section~\ref{sec:collapsing} to make contact with existing results on gGSt in the literature. But even for more general structure groups, the partial gauge fixing and the residual gauge transformations arise at the same time order by order from \eqref{eqn:gauge}.

As $A$ plays the role of a connection, it does not transform covariantly and it is more instructive to study the anomalous part
\begin{equation}\label{eqn:DeltaA}
	\Delta A = \delta A - [ \xi, A]\,
\end{equation}
of its transformation instead of the full transformation $\delta A$. In terms of this new quantity, \eqref{eqn:gauge} becomes
\begin{equation}\label{eqn:gaugeDelta}
	\Delta A^{(n)} + \delta E^{(n)} E^{-1}  - \widetilde{\xi}^{(n)} = \left. - D \xi - \mathcal{A}^{-1} \Delta \mathcal{A} \right|^{(<n)}\,.
\end{equation}
An advantage of this rewriting is that it becomes straightforward to implement generalized diffeomorphisms with the parameter $\xi_A$ by just extending $\Delta A$ to
\begin{equation}
	\Delta' A = \delta' A - [ \xi, A ] - \xi^B D_B A = \Delta A
\end{equation}
as
\begin{equation}
	\delta' A = \delta A + \xi^A D_A A\, .
\end{equation}
Therefore, $\Delta' A$ and $\Delta A$ constitute the same object, thereby we will focus our study only in the unprimed version. But as this substitution does not change the form of \eqref{eqn:gaugeDelta}, the results for the gauge fixing and the gGSt are still the same. This is not a coincidence, but rather a feature of the \PS construction, where generalized diffeomorphisms are a manifest symmetry. Consequently, we continue with $\xi_A = 0$, as it does not give rise to new structures.

To disentangle the left-hand side further, we keep in mind that $\widetilde{\xi}$ is chiral, while both $\Delta A^{(n)}$ and $\delta E^{(n)} E^{-1}$ for $n>0$ are mixed chiral. Moreover, $\delta E^{(n)} E^{-1}$ only contributes with $K_{A B}$ generators, while $\delta A^{(n)}$ can only contain the two different types of $R$-generators. After separating the two of them, we are thus left with
\begin{equation}
	\delta E^{(n)} E^{-1} = - \mathcal{A}^{-1} \Delta \mathcal{A} - [A, \widetilde{\xi}] \big|^{(<n)}_{K_-}\, ,
\end{equation}
where the subscript $K_-$ indicates that we are considering only the mixed chirality $K_{A B}$ part. Next, we eliminate the gauge fixed transformation parameter $\widetilde{\xi}$ from the right-hand side by using \eqref{eqn:gaugeDelta} to find
\begin{equation}
	\label{eqn:deltaEfull}
	\delta E^{(n)} E^{-1} = - [A, D\xi_+] \big|^{(<n)}_K - X^{(n)}\, ,
\end{equation}
with
\begin{equation}\label{eqn:X1}
	X = \mathcal{A}^{-1} \Delta \mathcal{A} + [A, \mathcal{A}^{-1} \Delta \mathcal{A}] \big|^{(<n)}_{K_-}\, .
\end{equation}
This is all that can be done without taking into account the explicit parametrization of $\mathcal{A}$. But after taking into account \eqref{eqn:sqrtCayley}, appendix~\ref{app:X} shows that an alternative way to write $X^{(n)}$ is
\begin{equation}\label{eqn:X2}
	X^{(n)} = \left. A \mathcal{A}^{-1} \Delta \mathcal{A} A \right|^{(<n)}_{K_-}\,.
\end{equation}
Again, we use \eqref{eqn:gaugeDelta} to eliminate $\mathcal{A}^{-1} \Delta \mathcal{A}$ in favor of
\begin{equation}
	X^{(n)} = - A \left( \delta E E^{-1} + D\xi \right) A  \big|^{(<n)}_{K_-}\,.
\end{equation}
Any non-trivial contribution to $X^{(n)}$ needs an $R_{\alpha\beta}$ generator in the brackets. However, $\delta E E^{-1}$ only gives rise to $K_{A B}$ generators, and $D\xi$ will only contain $R_\alpha^A$ generators. Therefore, we conclude $X^{(n)}=0$ and are left with the simple, all-order result
\begin{equation}\label{eqn:dEfinal}
	\boxed{\delta E E^{-1} = -  [A, D\xi_+ ] \big|_K}
\end{equation}
for the gGSt, with the abbreviation $D\xi_+$ for the chiral part of $D\xi$ (while $D\xi_-$ is the mixed chiral contribution).

To obtain the final expression for the transformation $\delta E E^{-1}$ which only depends on the generalized frame $E$ and its derivatives, one has to solve the torsion constraints to eliminate $A$. We will come back to this point in section~\ref{sec:collapsing}.

\section{Gauge algebra}\label{sec:alg}
As the gGSt \eqref{eqn:dEfinal} arises in our approach from a partial gauge fixing, it is not directly obvious that these transformations will close. At the leading orders this closure has been shown \cite{Baron:2018lve,Baron:2020xel}, but will it also hold to all orders? To answer this question, we compute the commutator of two transformations. More precisely, we apply a second transformation to \eqref{eqn:gaugetr}, resulting in
\begin{equation}
	\begin{aligned}
		\label{eqn: deltagauge}
		 & 2 \delta_{[2}\left(\delta_{1]}E E^{-1}\right) + 2 \delta_{[2}\left(\mathcal{A}^{-1}\delta_{1]}\mathcal{A}\right) =                  \\
		 & - 2 \delta_{[2}\left(D\xi_{{1]}}\right) - 2 \delta_{[2}\xi_{1]R} + 2\delta_{[2}\left(\mathcal{A}^{-1}\xi_{1]}\mathcal{A}\right) \,.
	\end{aligned}
\end{equation}
After distributing all variations there are only
\begin{equation}
	\delta_{[2} \delta_{1]}A^{(n)}\,, \qquad  \delta_{[2}\delta_{1]}E^{(n)} E^{-1} \,, \quad \text{and} \quad \delta_{[2} \widetilde{\xi}^{(n)}_{1]}
\end{equation}
as unknown terms left. They have to be fixed recursively, in the same way as we already did for the gauge transformation in \eqref{eqn:gauge}. There, everything worked because we could separate the terms on the left-hand side by generators. This is also possible for $\delta_{[2} \delta_{1]}A^{(n)}$; like $\delta A^{(n)}$, it has only contributions from mixed chiral $R^A_{\alpha}$ and $R_{\alpha\beta}$ generators. Moreover, $\delta_{[2} \widetilde{\xi}^{(n)}_{1]}$ is purely chiral like $\widetilde{\xi}^{(n)}$. Therefore, one would expect that $\delta_{[2}\delta_{1]}E^{(n)} E^{-1}$ is formed by chiral $K_{A B}$ generators at the zeroth order, followed by their mixed chiral counterparts at higher orders. For convenience, let us denote these contributions by $K_\pm$. Here, we adopt the notation that the upper sign is for the zeroth order and the rest for all higher ones. Clearly this is true for $\delta_{[2} ( \delta_{1]} E E^{-1} )$, and also for $\delta E E^{-1}$ directly. However, we find that
\begin{equation}
	\delta_{[2} \delta_{1]} E E^{-1} = \delta_{[2} ( \delta_{1]} E E^{-1} ) - \tfrac12 [ \delta_2 E E^{-1}, \delta_1 E E^{-1} ]
\end{equation}
cannot just be written in terms of generators in $K_\pm$ because of the last term on the right-hand side, originating from the transformation of $E^{-1}$. It includes both chiral and mixed chiral $K_{A B}$ generators, enabling us to write
\begin{equation}
	\begin{aligned}
		\relax[\delta_2 E E^{-1}, \delta_1 E E^{-1}] & = [\delta_2 E E^{-1}, \delta_1 E E^{-1}] \Big|_{K_{\mp}}     \\
		                                             & + [\delta_2 E E^{-1}, \delta_1 E E^{-1}] \Big|_{K_{\pm}}\, ,
	\end{aligned}
\end{equation}
where ${K_\mp}$ denotes the complement of ${K_\pm}$, namely mixed chiral terms in the zeroth order (although vanishing) and chiral terms for higher ones.

Having these at hand, we can now calculate $\delta_{[2} \delta_{1]} E E^{-1}$ directly from \eqref{eqn: deltagauge}. Applying all variations and taking into account that $D\xi$ transforms as
\begin{equation}
	\delta \left(D\xi \right) = [\delta E E^{-1}, D\xi] + D(\delta\xi)\,,
\end{equation}
we are left with
\begin{equation}\label{eqn: ddEinit}
	\begin{aligned}
		2\delta_{[2} & \delta_{1]} E E^{-1} = - 2 [\delta_{[2}E E^{-1}, D\xi_{1]}] - D\left(2\delta_{[2}\xi_{1]}\right)                                                         \\
		             & - 2\delta_{[2}\xi_{1]R} + 2\mathcal{A}^{-1} \delta_{[2}\xi_{1]}\mathcal{A}+ [\mathcal{A}^{-1}\delta_2 \mathcal{A}, \mathcal{A}^{-1}\delta_1 \mathcal{A}] \\
		             & - 2 [\mathcal{A}^{-1}\xi_{[2}\mathcal{A}, \mathcal{A}^{-1}\delta_{1]}\mathcal{A}]- 2 \mathcal{A}^{-1}\delta_{[2}\delta_{1]}\mathcal{A} \Big|_{K_\pm}     \\
		             & -  [\delta_2 E E^{-1}, \delta_1 E E^{-1}]\, .
	\end{aligned}
\end{equation}
Next, we eliminate $\mathcal{A}^{-1} \delta \mathcal{A}$ by using \eqref{eqn:gaugetr} again and thereby obtain
\begin{equation}\label{eqn:ddA+ddE}
	\begin{aligned}
		2\delta_{[2} & \delta_{1]}E E^{-1} = [D\xi_2, D\xi_1]                                                                                             \\
		             & - D\left(2\delta_{[2}\xi_{1]}\right) + 2[D\xi_{[2}, \xi_{1]R}]                                                                     \\
		             & - 2\delta_{[2}\xi_{1]R} + [\xi_{2R}, \xi_{1R}]                                                                                     \\
		             & + 2\mathcal{A}^{-1} \delta_{[2}\xi_{1]}\mathcal{A} - \mathcal{A}^{-1}[\xi_2, \xi_1]\mathcal{A}                                     \\
		             & - 2 \mathcal{A}^{-1}\delta_{[2}\delta_{1]}\mathcal{A} \Big|_{K_{\pm}} - [\delta_2 E E^{-1}, \delta_1 E E^{-1}] \Big|_{K_{\mp}} \,.
	\end{aligned}
\end{equation}
Looking closer at the second, third and fourth line, we identify the combination
\begin{equation}
	\label{eqn: xi21}
	\boxed{\xi_{21} = -[\xi_2, \xi_1] + 2\delta_{[2}\xi_{1]} = \xi_{21}^\alpha \tau_\alpha } \,,
\end{equation}
which allows us to simplify this relation considerably (more details about the required relations are given in appendix~\ref{app:X}) to
\begin{equation}\label{eqn: ddE}
	\begin{aligned}
		2\delta_{[2} & \delta_{1]}E E^{-1} = [D\xi_2, D\xi_1]                                                                                            \\
		             & - D\xi_{21} - \xi_{21R} - 2 \mathcal{A}^{-1}\delta_{[2}\delta_{1]}\mathcal{A}  +\mathcal{A}^{-1}\xi_{21}\mathcal{A} \Big|_{K_\pm} \\
		             & - [\delta_2 E E^{-1}, \delta_1 E E^{-1}] \Big|_{K_{\mp}} \,.
	\end{aligned}
\end{equation}
As before, we use the shorthand notation $\xi_{21 R} = \xi^\alpha_{21} R_\alpha$ and $\xi_{21} = \widetilde{\xi}_{21} + \xi_{21R}$ here. Also, we will drop the $K$ subscript from the last term, as it can only give rise to $K_{A B}$ contributions.

In the same vein as we proceeded from \eqref{eqn: ddEinit} to \eqref{eqn: ddE}, we can also treat the full expression \eqref{eqn:gauge} with an additional variation to obtain
\begin{equation}
	\label{eqn:ddgaugen}
	\begin{aligned}
		2\delta_{[2} & \delta_{1]}A^{(n)} + 2\delta_{[2}\delta_{1]}E^{(n)} E^{-1}-\widetilde{\xi}^{(n)}_{21} = [D\xi_2, D\xi_1]                       \\
		             & -\left. D\xi_{21} - 2 \mathcal{A}^{-1}\delta_{[2}\delta_{1]}\mathcal{A}+\mathcal{A}^{-1}\xi_{21}\mathcal{A} \right|^{(<n)} \,.
	\end{aligned}
\end{equation}
Because here all terms contribute, no projectors are needed. The last line looks remarkably close to the right-hand side of the transformation \eqref{eqn:gauge}. In order to get a complete match, first note that $\mathcal{A}^{-1}\delta_{[2}\delta_{1]}\mathcal{A}$ produces two different types of terms: First, we have terms where the transformations $\delta_1$ and $\delta_2$ act on two different $A$'s in the expansion of $\mathcal{A}$. But there are also terms where both of them act on the same $A$. Let us denote the latter by $\mathcal{A}^{-1} ( \delta_{[2}\delta_{1]} ) \mathcal{A}$. Due to the anti-symmetrization in $2$ and $1$, all terms with only one transformation on an $A$ cancel and one finds that
\begin{equation}
	\mathcal{A}^{-1} ( \delta_{[2}\delta_{1]} ) \mathcal{A} = \mathcal{A}^{-1} \delta_{[2}\delta_{1]} \mathcal{A}
\end{equation}
holds. Now, it is not hard to find the solution
\begin{equation}
	2\delta_{[2}\delta_{1]} A =  \delta_{21} A
\end{equation}
of \eqref{eqn:ddgaugen} for $2\delta_{[2}\delta_{1]} A$. Based on it, we can rewrite \eqref{eqn:ddA+ddE} as
\begin{equation}\label{eqn:ddEEinv}
	\begin{aligned}
		2\delta_{[2}\delta_{1]}  E E^{-1} = & [D\xi_{2}, D\xi_{1}] + \left.\delta_{21}E E^{-1}\right|_{K_{\pm}}       \\
		                                    & - \left.[ \delta_2 E E^{-1}, \delta_1 E E^{-1} ] \right|_{\mp}  \,.
	\end{aligned}
\end{equation}
In line with \eqref{eqn:gauge}, we want to have a $\delta_{[2}\delta_{1]}E^{(n)} E^{-1}$ that is purely mixed chiral for $n > 0$. To do so, we need to shift $[ \delta_2 E E^{-1}, \delta_1 E E^{-1} ]_{\mp}$ to a fully chiral object and the natural choice is $\widetilde{\xi}^{(n)}$. Indeed, plugging \eqref{eqn:ddEEinv} in \eqref{eqn:ddgaugen} we obtain
\begin{equation}
	\label{eqn:ddgaugefinal}
	\begin{aligned}
		\delta_{21} & A^{(n)} + \delta_{21}E^{(n)} E^{-1}-\widetilde{\xi}^{(n)}_{21} =                                         \\
		            & [D\xi_{2+}, D\xi_{1+}] + [D\xi_{2-}, D\xi_{1-}] + [ \delta_2 E E^{-1}, \delta_1 E E^{-1} ] \big|_{\mp} \\
		            & -\left.\mathcal{A}^{-1}\delta_{21}\mathcal{A}+\mathcal{A}^{-1}\xi_{21}\mathcal{A} \right|^{(<n)} \,,
	\end{aligned}
\end{equation}
from which we can identify
\begin{equation}
	\label{eqn:xi21t}
	\begin{aligned}
		\widetilde{\xi}_{21}^{(n)} = & \underline{- [D\xi_{2+}, D\xi_{1+}] - [D\xi_{2-}, D\xi_{1-}]}                    \\
		                             & \underline{-\left.[ \delta_2 E E^{-1}, \delta_1 E E^{-1} ]_{\mp}\right|^{(< n)}} \\
		                             & + \widetilde{\xi}^{(n)}\left(\xi_{21}^{(< n)}\right) \,,
	\end{aligned}
\end{equation}
while \eqref{eqn:ddEEinv} also reduces to
\begin{equation}\label{eqn:ddEEfinal}
	\begin{aligned}
		2\delta_{[2}\delta_{1]}E^{(n)} E^{-1} = & \phantom{-} \underline{\left.2[D\xi_{[2+}, D\xi_{1]-}]\right|^{(< n)}} \\
		                                        & + \delta_{21}E^{(n)}E^{-1} \,.
	\end{aligned}
\end{equation}
As a remark, \eqref{eqn:xi21t} should not be seen as a contradiction to \eqref{eqn: xi21} but rather as a reexpression of the latter's tilded version. The correspondence between the two is straightforward using \eqref{eqn:tildexi1}-\eqref{eqn:tildexi4} as examples.

Without the underlined terms, closure of the algebra with the new parameter $\xi_{21}$ would be immediately obvious. But they are there, and thus we have to explain their implications. Let us look first at \eqref{eqn:xi21t}.  Due to the section condition the underlined terms in the first line can only contribute with a $K_{A B}$ generator. Hence, all three affect the double Lorentz parameter. Originally, it has been set to $\widetilde{\xi}^{(0)} = \Lambda^{(0)}$, but now we are rather dealing with
\begin{equation}\label{eqn:Lambda21}
	\boxed{
		\begin{aligned}
			\Lambda^{(n)}_{21} = & - [D\xi_{2+}, D\xi_{1+}] - [D\xi_{2-}, D\xi_{1-}]                    \\
			                     & -\left.[ \delta_2 E E^{-1}, \delta_1 E E^{-1} ]_{\mp}\right|^{(< n)}
		\end{aligned}}
\end{equation}
for $n>0$ in addition to $\Lambda^{(0)}_{21} = - [\Lambda^{(0)}_2, \Lambda^{(0)}_1]$. This allows us to write
\begin{equation}
	\widetilde{\xi}_{21} = \widetilde{\xi}\left( \Lambda_{21}\right)\,.
\end{equation}
For the transformation of the generalized frame \eqref{eqn:ddEEfinal}, we have to remember that the generalized frame also transforms under generalized diffeomorphisms. Therefore instead of $\delta E E^{-1}$, which just covers double Lorentz and gGSt, we rather should consider
\begin{equation}
	\delta'_{21} E E^{-1} = \delta_{21} E E^{-1} + \left. 2 ( D^B \xi_{21}^A + \xi_{21}^C F_C{}^{AB} ) K_{AB} \right|_-\, .
\end{equation}
The underlined term in \eqref{eqn:ddEEfinal} exactly arises from the second term in this transformation for the generalized diffeomorphism parameter
\begin{equation}
	\label{eqn:xi21A}
	\boxed{\xi^A_{21} = -\xi^{\alpha}_{[2}D^A \xi_{1]\alpha}}\,,
\end{equation}
allowing us eventually to write
\begin{equation}
	2\delta_{[2}\delta_{1]}E^{(n)} E^{-1} = \delta'_{21} E^{(n)} E^{-1}
\end{equation}
and thereby completing the proof of the gGSt's closure.

\section{Torsion constraints}\label{sec:collapsing}
To compare the leading orders of our gGSt \eqref{eqn:dEfinal} and its gauge algebra with results in the literature, we have to finally solve the torsion constraints to eliminate $A$. All required steps are demonstrated in detail in \cite{Gitsis:2024gfb}. Here, we just want to emphasize the subtlety that the torsion constraints also contain a second type of generators, $t_{\alpha}$, which are related to $\tau_{\alpha}$ by the similarity transformation
\begin{equation}
	t_{\alpha} = S_{\alpha}{}^{\beta}\tau_{\beta} \,.
\end{equation}
Tensors in this alternative basis for the Lie algebra of $\GS$ are marked with a tilde, for example
\begin{equation}
	\widetilde\xi = \xi^\alpha t_\alpha
	\qquad \text{and} \qquad
	A_A{}^\beta t_\beta = \widetilde{A}_A{}^\beta \tau_\beta\,.
\end{equation}
One encounters the same challenge for the partial gauge fixing: In \eqref{eqn:tildexi1}-\eqref{eqn:tildexi4}, we find $\widetilde{\xi}$ on the left-hand side, while the right-hand side only contains the untilded version $\xi$. Hence, similarity transformations have to be performed recursively. Rather than $S_\alpha{}^\beta$, which is relatively simple, its more complicated inverse $(S^{-1})_{\alpha}{}^{\beta}$ is needed. It arises from a sum of increasing powers of $f_{\alpha\beta}{}^{\gamma}$. As a result, many, although redundant, terms contribute to $\xi$ and $A$ and complicate the computations. We remedy this challenge by a procedure called ``collapsing towers'', a scheme to simplify this sum by sending the dimension of $\GS$ to infinity.

\subsection{Generalized Green-Schwarz for odd number of derivatives}
Expanding \eqref{eqn:dEfinal} in derivatives, it is straightforward to verify that for $n=0,2,4$ the same results arise as the ones presented in \cite{Gitsis:2024gfb}. This is not surprising, as the motivation for this letter is to extend the results there to all orders in $\alpha'$. However, $n$ is not restricted to only even integers and we know that at least for $n=1,3$, $\delta E^{(n)} E^{-1}$ has to vanish. As \eqref{eqn:gauge} immediately gives rise to
\begin{equation}
	\delta E^{(1)}E^{-1} = 0 \,,
\end{equation}
we find that our expectation is met for $n=1$. The situation, however, becomes more elusive when we go higher in derivatives.
For $n=3$, \eqref{eqn:dEfinal} reads
\begin{equation}
	\delta E^{(3)}E^{-1} = - [A^{(2)}, D\xi^{(0)}_+] - [A^{(1)}, D\xi^{(1)}_+]
\end{equation}
and after collapsing the towers, it takes the form
\begin{equation}
	\begin{aligned}
		\delta E^{(3)}E^{-1} \sim & (\widetilde{A}_{A}^{(2)\alpha_2}D_B \widetilde{\xi}_+^{(0)\beta_1}\kappa_{\alpha_2\beta_1}             \\
		                          & + \widetilde{A}_{A}^{(1)\alpha_1}D_B \widetilde{\xi}_+^{(1)\beta_2}\kappa_{\alpha_1\beta_2})K^{A B}\,,
	\end{aligned}
\end{equation}
where the Greek indices have been further decomposed as explained in \cite{Gitsis:2024gfb}. Here, $\sim$ denotes modulo an overall coefficient which is irrelevant for our discussion. However, due to the fact that $\kappa_{\alpha_i\beta_j}$ is block diagonal,
\begin{equation}
	\kappa_{\alpha_i \beta_j} = 0 \qquad  \text{for} \qquad i \neq j \,,
\end{equation}
$\delta E^{(3)}E^{-1}$ eventually vanishes. It should be stressed that as it vanishes only after imposing the torsion constraints and collapsing the towers, we keep it in all the equations, for example in appendix~\ref{app:params}.

\subsection{Gauge algebra results}
To make contact with the results for the algebra created by gauge transformations in \cite{Baron:2018lve,Baron:2020xel}, we only need to consider \eqref{eqn:Lambda21} for the respective order. After imposing the torsion constraints and collapsing the towers, we obtain
\begin{align}
	\Lambda_{21}^{(0)\aR\bR} & = -2 \Lambda^{\aR \Cr}_{[2}\Lambda_{1]\Cr}{}^{\bR}\,,                                                                                                                                                                                         \\
	\Lambda_{21}^{(2)\aR\bR} & = -a D^{\underline{a}}\Lambda_{[2}^{\underline{c}\underline{d}}D^{\underline{b}}\Lambda_{1]\underline{c}\underline{d}} + b D^{\underline{a}}\Lambda_{[2}^{\overline{c}\overline{d}}D^{\underline{b}}\Lambda_{1]\overline{c}\overline{d}} \, ,
\end{align}
\begin{widetext}
	\begin{equation}
		\begin{aligned}
			\Lambda_{21}^{(4)\aR\bR} = & \phantom{-} a b \left[F^{\gl}{}_{\Cr\dr} D^{[\aR}\Lambda_{[2}^{\Cr\dr}F^{\bR]}{}_{\el\fl}D_{\gl}\Lambda_{1]}^{\el\fl} - D^{[\aR}\Lambda_{[2\el\fl}D^{\bR]}\left(F^{\el}{}_{\Cr\dr}D^{\fl}\Lambda_{1]}^{\Cr\dr}\right)- D^{[\aR}\Lambda_{[2\er\fr}D^{\bR]}\left(F^{\er}{}_{\cl\dl}D^{\fr}\Lambda_{1]}^{\cl\dl}\right)\right] \\
			                           & + a^2 \left[\frac{1}{2}D^{\aR}\Lambda_{[2}^{\Cr\dr}D^{\bR}\Lambda_{1]}^{\er\fr}F^{\gl}{}_{\Cr\dr}F_{\gl\er\fr} + D^{\aR}D^{\er}\Lambda_{[2}^{\Cr\dr}D^{\bR}D_{\er}\Lambda_{1]\Cr\dr} + 2 D^{[\aR}\left(D^{\cl}\Lambda_{[2}^{\er\dr}F_{\cl\dr}{}^{\fr}\right)D^{\bR]}\Lambda_{1]\er\fr} \right]                              \\
			                           & + b^2 \left[\frac{1}{2}D^{\cl}\Lambda_{[2}^{\el\fl}D_{\cl}\Lambda_{1]}^{\dl\gl}F^{\aR}{}_{\el\fl}F^{\bR}{}_{\dl\gl} + D^{\aR}D^{\el}\Lambda_{[2}^{\cl\dl}D^{\bR}D_{\el}\Lambda_{1]\cl\dl} + 2 D^{[\aR}\left(D^{\Cr}\Lambda_{[2}^{\el\dl}F_{\Cr\dl}{}^{\fl}\right)D^{\bR]}\Lambda_{1]\el\fl}\right]\, .
		\end{aligned}
	\end{equation}
\end{widetext}
in the leading orders. These expressions match the results from the gBdRi \footnote{Note that there initially has been a deviation between our results and \cite{Baron:2018lve,Baron:2020xel}. Thanks to the help of Diego Marques and Walter Baron who kindly agreed to revisit their computation, this problem has been solved.} \cite{Baron:2018lve,Baron:2020xel}. We do not present $\Lambda_{21}^{\aL\bL}$ here explicitly because it arises from $\Lambda_{21}^{\aR\bR}$ after flipping all chiralities and exchanging $a \longleftrightarrow -b$. Moreover, up to four derivatives, the diffeomorphism parameter \eqref{eqn:xi21A} reads
\begin{align}
	 & \xi_{21}^{(1)A}   = \frac{a}{2}\Lambda_{[2}^{\underline{b}\underline{c}}D^A\Lambda_{1]\underline{b}\underline{c}} - \frac{b}{2}\Lambda_{[2}^{\overline{b}\overline{c}}D^A\Lambda_{1]\overline{b}\overline{c}}\,,                                                                                                                                                                                                          \\
	 & \xi^{(3)A}_{21} =                                                                                                                                                                                                -\frac{a b}{2}\left(D^A \Lambda_{[2}^{\er\fr} D_{\fr}\Lambda_{1]\cl \dl} F_{\er}{}^{\cl\dl} +D^A \Lambda_{[2}^{\el\fl} D_{\fl}\Lambda_{1]\Cr \dr} F_{\el}{}^{\Cr\dr} \right)                   \nonumber \\
	 & + a^2 \left[D^A \Lambda_{[2}^{\er\fr}D^{\cl}\Lambda_{1]\er}{}^{\dr}F_{\cl\dr\fr} + \frac{1}{2}D^A\left(D^{\Cr}\Lambda_{[2}^{\er\fr}\right)D_{\Cr}\Lambda_{1]\er\fr}\right]                     \nonumber                                                                                                                                                                                                                  \\
	 & + b^2 \left[D^A \Lambda_{[2}^{\el\fl}D^{\Cr}\Lambda_{1]\el}{}^{\dl}F_{\Cr\dl\fl} + \frac{1}{2}D^A\left(D^{\cl}\Lambda_{[2}^{\el\fl}\right)D_{\cl}\Lambda_{1]\el\fl}\right]\, ,
\end{align}
which again matches \cite{Baron:2018lve,Baron:2020xel} after setting $\xi^A = 0$.

\section{Conclusions}
We succeeded in computing gGSt to all orders in $\alpha'$. The key to obtain the simple transformation law \eqref{eqn:dEfinal} is to decouple torsion constraints from the partial gauge fixing by following the strategy we developed in \cite{Gitsis:2024gfb}. Moreover, we identified a preferred parameterization of the field $\mathcal{A}$ in \eqref{eqn:sqrtCayley}. Although they do not qualitatively change any of our conclusions, alternative choices of $\mathcal{A}$ will lead to much more complicated expressions. Furthermore, it is possible to show that our all-order gGSt close after combining them with generalized diffeomorphisms. The resulting transformations are parameterized by \eqref{eqn:Lambda21} and \eqref{eqn:xi21A}. As a consistency check, we also computed the final version of the transformations and their algebra just in terms of the generalized frame $E_A{}^I$ and its flat derivatives. As expected, our results match with the gBdRi \cite{Baron:2018lve,Baron:2020xel}.

As pointed out by \cite{Hronek:2020xxi,Hsia:2024kpi}, the O($d$,$d$) symmetry discussed here is not able to capture all $\alpha'$-corrections which arise in string theory. It only captures a subsector. Therefore a very important, but also complicated, problem is to understand if this symmetry can be extended/adapted such that all corrections which arise in string theory are accommodated. But even without such an extension, gGSt are definitely important to get a better handle on higher-derivative constructions for generalized dualities \cite{Hassler:2020tvz,Borsato:2020wwk,Codina:2020yma}. Another interesting direction originates from the fact that the partial gauge fixing \eqref{eqn:gaugefixing} and the torsion constraints we used here are not fixed uniquely. It has been demonstrated that an alternative realization of the torsion constraints produces rather simple results for gauge transformations and the action \cite{Hronek:2022dyr}. Thus, it would be interesting to see if there is a preferred choice that simplifies the final expressions written in terms of the metric and the $B$-field.

\begin{acknowledgments}
	We would like to thank Walter Baron and Diego Marques for very helpful correspondence and comments on the draft. AG and FH are supported by the SONATA BIS grant 2021/42/E/ST2/00304 from the National Science Center (NCN),	Poland.
\end{acknowledgments}

\appendix
\section{Identities}\label{app:X}
For any differential operator, like $\delta$ or $\nabla$, where we use here $\partial$ as a placeholder, we find the expansion
\begin{equation}
	\mathcal{A}^{-1} \partial \mathcal{A} = \sum\limits_{m=0}^\infty \sum\limits_{n=0}^\infty d_{m,n} A^m \partial A A^n\, ,
\end{equation}
with
\begin{equation}
	d_{m,n} = \sum\limits_{l=0}^m (-1)^l d_{m+n+1-l}\,.
\end{equation}
By using this expansion, the equivalence between \eqref{eqn:X1} and \eqref{eqn:X2} is implied by
\begin{equation}
	d_{2m,2n} + d_{2m\pm 1,2n} - d_{2m,2n\pm 1} - d_{2m\pm 1,2n\pm 1} = 0\,,
\end{equation}
which holds due to the equivalence between even and odd coefficients given in \eqref{eqn:dn}. Note that the explicit values of the coefficients are not relevant here.

Furthermore, to see more explicitly how to get from \eqref{eqn:ddA+ddE} to \eqref{eqn: ddE}, first note that
\begin{equation}
	[\xi_{2R}, \xi_{1R}] = \xi_{2}^{\alpha}\xi_{1}^{\beta}[R_{\alpha}, R_{\beta}] = -\xi_2^{\alpha}\xi_1^{\beta}f_{\alpha\beta}{}^{\gamma}R_{\gamma}\,
\end{equation}
holds and therefore substitute
\begin{align}
	-2\delta_{[2}\xi_{1]R} + [\xi_{2R}, \xi_{1R}] & = -\left(2 \delta_{[2}\xi_{1]}^{\gamma} + \xi_2^{\alpha}\xi_1^{\beta}f_{\alpha\beta}{}^{\gamma} \right)R_{\gamma} \nonumber \\
	                                              & = - \xi_{21R}
\end{align}
in the third line of \eqref{eqn:ddA+ddE}.  In the same vein, one finds
\begin{equation}
	- D\left(2\delta_{[2}\xi_{1]}\right) + 2[D\xi_{[2}, \xi_{1]R}] = - D\xi_{21}
\end{equation}
for the second line.

\section{Gauge fixed parameters}\label{app:params}
Here, we provide expressions for the gauge parameters $\widetilde{\xi}$ for reference. We go up to four derivatives as the equations become progressively lengthier and we get no further insights.  For $\widetilde{\xi}$, we need to project the right-hand side of \eqref{eqn:gauge} in (anti-)chiral terms. Doing so, we find
\begin{widetext}
	\begin{align}\label{eqn:tildexi1}
		\widetilde{\xi}^{(1)} = & D\xi^{(0)}_{+}\, ,                                                                                                                                                      \\
		\widetilde{\xi}^{(2)} = & D\xi^{(1)}_{+} + \frac{1}{2}[A^{(1)}, D\xi^{(0)}_-]\, ,                                                                                                                 \\
		\widetilde{\xi}^{(3)} = & D\xi^{(2)}_{+} + \frac{1}{2}[A^{(2)}, D\xi^{(0)}_-] + \frac{1}{2}[A^{(1)}, D\xi^{(1)}_- + \delta E^{(2)}E^{-1}]\, ,                                                     \\
		\widetilde{\xi}^{(4)} = & D\xi^{(3)}_{+} + \frac{1}{2}\left[A^{(3)} + \frac{1}{3}A^{(1)}A^{(1)}A^{(1)}, D\xi^{(0)}_-\right] + \frac{1}{2}[A^{(2)}, D\xi^{(1)}_- + \delta E^{(2)}E^{-1}] \nonumber \\
		                        & + \frac{1}{2}[A^{(1)}, D\xi^{(2)}_- + \delta E^{(3)}E^{-1}]-\frac{1}{24}[A^{(1)}, [A^{(1)}, [A^{(1)}, D\xi^{(0)}_-]]]\label{eqn:tildexi4} \,.
	\end{align}
\end{widetext}

\bibliography{literature.bib}

\end{document}